\documentclass[preprint,aps,showpacs,nofootinbib,preprintnumbers,amsmath,amssymb,superscriptaddress]{revtex4-2}
\usepackage{amsmath,amssymb,amsthm,mathrsfs}
\usepackage{epsfig}
\usepackage{graphicx}
\usepackage[usenames,dvipsnames]{color}
\usepackage{subfigure}
\usepackage{slashed}
\usepackage{comment}
\usepackage[normalem]{ulem}
\usepackage{tikz}
\usepackage[compat=1.1.0]{tikz-feynman}
\usepackage[colorlinks,citecolor=blue]{hyperref}
\usepackage{color}

\usepackage{multirow}
\usepackage{tikz-feynman}
\tikzfeynmanset{compat=1.1.0}

\usepackage{tikzsymbols}

\usepackage{xcolor}
\usepackage{soul}

\begin{document}

\title{Cosmic Birefringence from Neutrino and Dark Matter Asymmetries}
\author{Ren-Peng Zhou\footnote{zhourenpeng21@mails.ucas.ac.cn}}
\affiliation{School of Fundamental Physics and Mathematical Sciences, Hangzhou Institute for Advanced Study, UCAS, Hangzhou 310024, China}
\affiliation{University of Chinese Academy of Sciences (UCAS), Beijing 100049, China}
\author{Da Huang\footnote{dahuang@bao.ac.cn}}
\affiliation{National Astronomical Observatories, CAS, Beijing, 100012, China}
\affiliation{School of Fundamental Physics and Mathematical Sciences, Hangzhou Institute for Advanced Study, UCAS, Hangzhou 310024, China}
\affiliation{University of Chinese Academy of Sciences (UCAS), Beijing 100049, China}
\affiliation{International Centre for Theoretical Physics Asia-Pacific, Beijing/Hangzhou, China}
\author{Chao-Qiang Geng\footnote{cqgeng@ucas.ac.cn}}
\affiliation{School of Fundamental Physics and Mathematical Sciences, Hangzhou Institute for Advanced Study, UCAS, Hangzhou 310024, China}
\affiliation{University of Chinese Academy of Sciences (UCAS), Beijing 100049, China}
\affiliation{International Centre for Theoretical Physics Asia-Pacific, Beijing/Hangzhou, China}

\begin{abstract}
In light of the recent measurement of the nonzero Cosmic Microwave Background (CMB) polarization rotation angle from the Planck 2018 data, we explore the possibility that such a cosmic birefringence effect is induced by coupling a fermionic current with photons via a Chern-Simons-like term.
We begin our discussion by rederiving the general formulae of the cosmic birefringence angle with correcting a mistake in the previous study. 
We then identify the fermions in the current as the left-handed electron neutrinos and asymmetric dark matter (ADM) particles, since the rotation angle is sourced by the number density difference between particles and antiparticles. For the electron neutrino case, with the value of the degeneracy parameter $\xi_{\nu_e}$ recently measured by the EMPRESS  survey, we find a large parameter space which can explain the CMB photon polarization rotations. On the other hand, for the ADM solution, we consider two benchmark cases with $M_\chi = 5$~GeV and 5~keV. The former is the natural value of the ADM mass if the observed ADM and baryon asymmetry in the Universe are produced by the same mechanism, while the latter provides a warm DM candidate. In addition, we  explore the experimental constraints from the CMB power spectra and the DM direct detections. 

\end{abstract}

\maketitle
\newpage
\section{Introduction}
\noindent Cosmic birefringence is a remarkable parity-violating phenomenon in which the plane of a linearly polarized photon rotates along its propagation path in the astronomical scale, which is caused by the small distinction in the phase velocities between the left- and right-handed photon polarizations~\cite{Carroll:1989vb,Carroll:1991zs,Harari:1992ea,Carroll:1998zi} (see {\it e.g.} Ref.~\cite{Komatsu:2022nvu} for a recent review on this issue). This effect can be imprinted in the Cosmic Microwave Background (CMB) polarization data as the parity-odd cross correlation between  $E$ and $B$ modes~\cite{Lue:1998mq}. Recently, the analysis of the CMB data from the {\it Planck} public release 4 (PR4) has shown a tantalizing nonzero value of the cosmic birefringence angle $ \Delta \alpha = 0.30\pm 0.11$~deg\footnote{We hereinafter denote the CMB cosmic birefringence angle with $\Delta \alpha$, rather than $\beta$ in the convention used in other literature~\cite{Diego-Palazuelos:2022dsq}.} at 68\% confidence level (CL)~\cite{Diego-Palazuelos:2022dsq}. By taking into account the Milky Way foreground $EB$ cross correlations, the CMB photon rotation angle is improved to be $\Delta \alpha = 0.36 \pm 0.11$~deg at 68\% CL, which indicates that the statistical significance exceeds $3\sigma$. Note that the new result given in Ref.~\cite{Diego-Palazuelos:2022dsq} is an update of the measurement of the isotropic birefringence angle $\Delta\alpha = 0.35\pm 0.14$~deg at 68\% CL from the {\it Planck} PR3 data~\cite{Minami:2020odp}, in which a new technique has been proposed to address the long-standing degeneracy problem of a miscalibration angle of polarimeters~\cite{QUaD:2008ado,WMAP:2010qai}. Later, the measurement of the birefringence angle $\Delta \alpha$ has been improved by including the high-frequency instrumental data~\cite{Eskilt:2022wav} and the {\it WMAP} data~\cite{Eskilt:2022cff}. If the above nonzero birefringence angle is confirmed in the future, it would provide us a novel evidence for new physics beyond the Standard Model (SM). In the literature, a typical explanation of this remarkable parity-violating effect usually involves the existence of a pseudoscalar axion-like field~\cite{Fujita:2020ecn,Takahashi:2020tqv,Fung:2021wbz,Choi:2021aze,Obata:2021nql,Gasparotto:2022uqo,Carroll:1998zi,Finelli:2008jv,Arvanitaki:2009fg,Panda:2010uq,Fedderke:2019ajk,Fujita:2020aqt,Mehta:2021pwf,Nakagawa:2021nme,Yin:2021kmx,Lin:2022niw,Murai:2022zur,Greco:2022xwj,Gonzalez:2022mcx,Galaverni:2023zhv,Kitajima:2022jzz,Cai:2022zad,Jain:2022jrp}, which can play the role of dark energy or dark matter (DM) in the Universe (see {\it e.g.} Refs.~\cite{Marsh:2015xka,Ferreira:2020fam} for recent reviews and references therein). By coupling the axion-like field to the photon Chern-Simons (CS) term~\cite{Ni:1977zz,Turner:1987bw}, the two circularly-polarized photon modes can be distinguished due to their different phase velocities, which can source the cosmic birefringence in the CMB data. 

In this work, we would like to study an alternative origin of the observed cosmic birefringence~\cite{Geng:2007va,Ho:2010aq}, which generates this parity-violating effect by allowing photons to couple to a fermionic current in a CS-like manner. In Refs.~\cite{Geng:2007va,Ho:2010aq}, such a mechanism is thoroughly explored by identifying the fermion particles in the current as the active neutrinos in the SM. However, as shown below, the analysis in Ref.~\cite{Geng:2007va} contained some mistake in the final results. In order to clarify the possible origin of the mistake, we first rederive the general formulae for the cosmic birefringence angle $\Delta \alpha$ for an arbitrary fermionic current. We then use the obtained expression of $\Delta \alpha$ to study two specific models in which the fermions are assumed to be the left-handed electron-type neutrinos and the DM particles, respectively. For the neutrino case, we would like to update the numerical results in Ref.~\cite{Geng:2007va} with the correct analytical formula of $\Delta \alpha$. As for the DM candidates, by noticing that it is the number density difference between particles and anti-particles that induces the photon birefringence, we are led to consider the asymmetric DM (ADM) scenario~\cite{Nussinov:1985xr,Kaplan:1991ah,Hooper:2004dc,Kaplan:2009ag} (For reviews of ADM models, see Refs~\cite{Petraki:2013wwa,Zurek:2013wia} and references therein), in which all the observed DM density in the Universe is solely composed of fermionic DM particles without its anti-particle counterparts. Instead of specifying the concrete mechanism for the ADM production, we would like to explore the phenomenology of this model at two benchmark points with the ADM mass to be $M_\chi = 5$~GeV and 5~keV. The former case is the natural ADM mass value if the DM and baryon relics are generated via the same mechanism in order to explain their cosmological mass density ratio, while the latter is a legitimate warm DM candidate~\cite{Bode:2000gq} which can help us to understand several small-scale structure problems~\cite{Klypin:1999uc}. We also take into account the experimental constraints on these two ADM cases, including the {\it Planck} CMB power spectra and the DM direct detection (DD) bounds.

The paper is organized as follows. In Sec.~\ref{derivation}, we rederive the general formulae for the cosmic birefringence angle $\Delta \alpha$ sourced by the Chern-Simons-like coupling of an arbitrary fermionic current to photons. Secs.~\ref{neutrino} and \ref{ADM} are dedicated to the phenomenological studies by identifying the fermions in the above current as the left-handed electron neutrinos and ADM particles, respectively. Finally, we summarize in Sec.~\ref{conclusion}. {In Appendix~\ref{SecLEP}, we provide a simple estimation of the LEP upper limits on the effective $\nu\bar{\nu}\gamma\gamma$ interaction, which might constrain the interpretation of the cosmic birefringence in terms of the $\nu_e$ asymmetry.}


\section{General Discussion of Cosmic Birefringence from a Fermion Current}\label{derivation}
In this section, we shall derive the formula of the isotropic birefringence angle induced by the coupling of a general fermion current $J_\mu$ to the photon Chern-Simons term~\cite{Geng:2007va,Ho:2010aq}. Let us begin our discussion by writing down the following Lagrangian~\cite{Geng:2007va}
\begin{align}\label{Lagrangian}
	\mathcal{L}=\mathcal{L}_{\rm EM}+\mathcal{L}_{\rm CS}
	=-\frac{1}{4}\sqrt{g}F_{\mu\nu}F^{\mu\nu}-\frac{1}{2}\sqrt{g}\frac{\beta}{M^2}J_\mu A_\nu \tilde{F}^{\mu\nu}\,,
\end{align}
where $g \equiv -{\rm det} (g_{\mu\nu})$ with $g_{\mu\nu}$ as the metric tensor of the spacetime, and $A_\mu$ denotes the electromagnetic field with its field strength as $F_{\mu\nu} = \partial_\mu A_\nu - \partial_\nu A_\mu$. We have also defined the dual field strength tensor as
\begin{eqnarray}
	\tilde{F}^{\mu \nu} \equiv \frac{1}{2} \epsilon^{\mu\nu\rho\sigma} F_{\rho\sigma}\,,
\end{eqnarray}
in which $\epsilon$ is the Levi-Civita tensor defined as $\epsilon^{\mu\nu\rho\sigma} \equiv g^{-1/2} e^{\mu\nu\rho\sigma}$ with $e^{\mu\nu\rho\sigma}$ as the antisymmetric symbol normalized to $e^{0123} = 1$. Since the coupling of $J_\mu$ to the photon Chern-Simons term is of six mass dimensions, we have introduced $M$ as the cutoff scale to balance the dimension with $\beta$ to be a constant of ${\cal O}(1)$. Note that the term ${\cal L}_{\rm CS}$ is not invariant under the electromagnetic $U(1)$ gauge transformation. However, as shown in Refs.~\cite{Geng:2007va,Ho:2010aq}, one could resort to the St\"{u}ckelberg mechanism or the anti-symmetric Kalb-Ramond field in order to maintain the gauge invariance.  

It is well-known that our universe is flat, homogeneous and isotropic, so that it can be described by the following Friedman-Lama\^{i}tre-Robertson-Walker {(FLRW)} metric 
\begin{eqnarray}\label{PhyFrame}
	ds^2 = -dt^2 + R^2(t) d\mathbf{x}^2\,,
\end{eqnarray}
where $t$ is the physical proper time while $\mathbf{x}$ denotes the spatial three-dimensional comoving coordinates with $R(t)$ being the scale factor. In this coordinate, the fermionic four-current is defined as $J_\mu = (J_t\,,\mathbf{J})$, where $\mathbf{J}$ is the fermion flux and $J_t$ is the number density difference between fermions and anti-fermions $J_t = \Delta n = n - \bar{n}$, with $n$($\bar{n}$) the number density of (anti-)fermions. In the present work, we only focus on the isotropic birefringence generated by $J_t$, and ignore the sub-leading anisotropic effects caused by the flux current $\mathbf{J}$. Thus, we fix $\mathbf{J} = 0$ for simplicity.  

We shall follow the procedure given in Refs.~\cite{Carroll:1989vb,Carroll:1991zs} to derive the birefringence angle induced by the current $J_\mu$. Firstly, we need to transform the coordinate metric into the following form
\begin{eqnarray}\label{ComovingFrame}
	ds^2 = R^2(\eta) (-d\eta^2 + d\mathbf{x}^2)\,,
\end{eqnarray}  
where $\eta$ is the conformal time with $d\eta = dt/R$. Also, {under the change of coordinate from Eq.~\eqref{PhyFrame} to Eq.~\eqref{ComovingFrame}, the fermion current $J_\mu$ is transformed according to {$J^\prime_\mu (x^\prime) = J_\nu(x) (\partial x^\nu/\partial x^{\prime\,\mu})$} where the quantities with (without) prime denote those defined in the FLRW (conformal) coordinate. But since the only non-zero component of $J_\mu$ is the temporal one, combined with the fact that the change of coordinate {does not mix} different components, we are led to $J^\prime_\mu (x^\prime) = \delta^t_\mu R(\eta) J_t(x)$, {\it i.e.}, $J_\eta (x^\prime) = R(\eta) J_t(x)$.}
By differentiating the Lagrangian in Eq.~(\ref{Lagrangian}) with respect to $A_\mu$, we can obtain the photon field equation 
\begin{eqnarray}\label{EOMF}
	\nabla_\mu F^{\mu\nu} = \frac{\beta}{M^2} J_\mu \tilde{F}^{\mu\nu}\,,
\end{eqnarray}
together with the Bianchi identities given by
\begin{eqnarray}\label{Bianchi}
	\nabla_\mu \tilde{F}^{\mu\nu} = 0\,.
\end{eqnarray}
In order to proceed, we can represent $F_{\mu\nu}$ and its dual $\tilde{F}_{\mu\nu}$ by the corresponding electric and magnetic fields, $\mathbf{E}$ and $\mathbf{B}$, as follows
\begin{eqnarray}\label{FieldStrength}
	F_{\mu\nu}={
		\begin{pmatrix}
			0 & -E_x & -E_y & -E_z\\
			E_x & 0 & B_z & -B_y\\
			E_y & -B_z & 0 & B_x\\
			E_z & B_y & -B_x & 0\\
	\end{pmatrix}}\,,\quad
	\tilde{F}_{\mu\nu}={
		\begin{pmatrix}
			0 & -B_x & -B_y & -B_z\\
			B_x & 0 & -E_z & E_y\\
			B_y & E_z & 0 & -E_x\\
			B_z & -E_y & E_x & 0
	\end{pmatrix}}\,.
\end{eqnarray}

As a result, the field equations in Eqs.~\eqref{EOMF} and \eqref{Bianchi} can be written as
\begin{eqnarray}\label{EOMEB}
	\frac{\partial \,\mathbf{E}}{\partial \eta}  -\nabla\times \mathbf{B} = \frac{\beta}{M^2} J_\eta \mathbf{B}  \,, \quad \nabla \cdot \mathbf{E} = 0\,,
\end{eqnarray}
and
\begin{eqnarray}\label{BianchiEB}
\frac{\partial\, \mathbf{B}}{\partial \eta}  + \nabla\times \mathbf{E} = 0\,,\quad \nabla\cdot \mathbf{B} = 0\,,
\end{eqnarray}
where $\nabla \cdot$ and $\nabla \times$ here denote the conventional differential operators in the three-dimensional Cartesian space. {Note that, by setting $\beta = 0$, the field equations in Eqs.~\eqref{EOMEB} and \eqref{BianchiEB} would be reduced to the standard Maxwell's equations in the vacuum. This can be understood in the fact that the electromagnetism is a conformally invariant theory given that the metric in Eq.~\eqref{ComovingFrame} is conformally equivalent to the flat Minkowski one.} By combining the equations in Eqs.~\eqref{EOMEB} and \eqref{BianchiEB} so as to eliminate $\mathbf{E}$, we can obtain 
\begin{eqnarray}\label{waveEOM}
	\frac{\partial^2\, \mathbf{B}}{\partial \eta^2}  - \nabla^2 \mathbf{B} = -\frac{\beta}{M^2} J_\eta \nabla\times \mathbf{B} \,.
\end{eqnarray} 

Now we consider the monochromatic wave solution to Eq.~\eqref{waveEOM} of the following form
\begin{eqnarray}\label{waveSoln}
	\mathbf{B} (\mathbf{x}, \eta) = e^{-i \mathbf{k}\cdot \mathbf{x}} \mathbf{B}(\eta)\,,
\end{eqnarray}
and assume that the wave propagates along the $z$ axis so that $\mathbf{k}\cdot \mathbf{x} = k z$. In addition, we define the two independent transverse-polarized waves in terms of their circular polarizations
\begin{eqnarray}\label{CircPol}
	F_{\pm} \equiv  B_{\pm} (\eta) =  B_x \pm i B_y \,,
\end{eqnarray} 
which can simplify the wave equation in Eq.~\eqref{waveEOM} into the following form
\begin{eqnarray}\label{EOMCirc}
	\frac{d^2 F_\pm}{d \eta^2} + \left(k^2 \pm \frac{\beta  k J_\eta}{M^2}\right) F_\pm =0\,.
\end{eqnarray}
By assuming that the fermion density $J_\eta$ evolves very slowly over the photon propagation, we can apply the WKB method to obtain the following approximated solution to Eq.~\eqref{EOMCirc}
\begin{eqnarray}
	F_\pm (\eta) = \exp\left[ik \int \left(1\pm \frac{\beta}{M^2} \frac{J_\eta}{k}  \right)^{1/2} d\eta \right] \,.
\end{eqnarray}  
Consequently, the solution to the original electromagnetic wave equation in Eq.~\eqref{waveEOM} is given by
\begin{eqnarray}
  	 B_\pm (z, \eta) = e^{-ikz} F_\pm (\eta) = e^{i\sigma_\pm}\,, 
\end{eqnarray} 
where the phase $\sigma_\pm$ is defined as
\begin{eqnarray}\label{phaseS}
	\sigma_\pm = k(\eta - z) \pm \frac{\beta}{2 M^2} \int J_\eta d\eta - \frac{\beta^2}{8 k M^4} \int J_\eta^2 d\eta + {\cal O}(k^{-2})\,.
\end{eqnarray}
{Note that we have made in Eq.~\eqref{phaseS} a Taylor expansion with respect to $\beta J_\eta/(k M^2)$ by assuming that this ratio is very small.} As a result, the birefringence polarization rotation angle induced by the Chern-Simons-like coupling ${\cal L}_{\rm CS}$ in Eq.~\eqref{Lagrangian} is given by \cite{Harari:1992ea,Carroll:1989vb}
\begin{eqnarray}\label{PolAngle}
	\Delta \alpha =  \frac{1}{2} \left(\sigma_+ - \sigma_-\right) \approx \frac{1}{2} \frac{\beta}{M^2} \int J_\eta d\eta = \frac{1}{2} \frac{\beta}{M^2} \int \Delta n\, dt\,,
\end{eqnarray}
where we have used the relations $J_\eta = R(\eta) \Delta n$ and $d\eta =dt/R$ in the last equality. Note that the formula for the angle of the polarization plane rotation in Eq.~\eqref{PolAngle} is different from Eq.~(10) in Ref.~\cite{Geng:2007va}, in which there was an extra scale factor $R$ in the denominator of the integrand. Such a distinction can be traced back to the mistreatment of the current density $J_t= \Delta n$ in deriving the wave equation in Eq.~\eqref{EOMEB} and the subsequent calculations, in which all quantities should defined in terms of conformal time so that $J_\eta$ should be employed. 

We would like to emphasize that, in deriving the general formula of the birefringence angle $\Delta \alpha$ in Eq.~\eqref{PolAngle}, we have not specified the origin of the fermionic current $J_\mu$. In the following two sections, we shall identify it as the current of left-handed electron neutrinos and fermionic ADM particles, both of which are of phenomenological importance. 



\section{Neutrino Current}\label{neutrino}
In this section, we shall identify $J_\mu$ as the active left-handed electron neutrino current as $J^{\nu_e}_\mu = \overline{({\nu}_{e})}_L \gamma_\mu (\nu_{e})_L$ in the SM. Such a case has already been explored in Refs.~\cite{Geng:2007va,Ho:2010aq}. However, as mentioned before, the improper dependence of the birefringence angle on the scale factor $R(t)$ given in Ref.~\cite{Geng:2007va} has made the analysis unreliable. Hence, here we would like to update the result of the neutrino-current-induced cosmic birefringence. According to our new $\Delta \alpha$ formula in Eq. \eqref{PolAngle}, the polarization angle rotation is given by
\begin{equation}\label{AngleNu}
    \Delta\alpha=\frac{1}{2}\frac{\beta}{M^2}\int J_\eta^{\nu_e} d\eta 
    =\frac{1}{2}\frac{\beta}{M^2}\int\Delta n_{\nu_e} dt\,,
\end{equation}
where $\Delta n_{\nu_e} \equiv n_{\nu_e} - n_{\bar{\nu}_e}$ is the density difference between neutrinos and anti-neutrinos in the coordinate defined in Eq.~\eqref{PhyFrame}. 
Note that the electron neutrino asymmetry is usually parameterized in the literature by the following parameter~\cite{Serpico:2005bc,Lesgourgues:2006nd}
\begin{align}\label{AsymParam}
\eta_{\nu_e}=\frac{\Delta n_{\nu_e}}{n_\gamma}=\frac{1}{12\zeta(3)}\left(\frac{T_\nu}{T_\gamma}\right)^3 \left(\pi^2\xi_{\nu_e}+{\xi_{\nu_e}}^3\right)\,,
\end{align}
{where $\zeta(z)$ is the Riemann function with $\zeta(3) \simeq 1.202$}, $n_{\gamma}$ is the number density of photons, $T_{\nu(\gamma)}$ is the temperature of neutrinos (photons), and $\xi_{\nu_e} =\mu_{\nu_e}/T_\nu$ is the degeneracy parameter with $\mu_{\nu_e}$ denoting the electron neutrino chemical potential, respectively. Based on the standard cosmological evolution, the temperature ratio between neutrinos and photons can be estimated as $(T_{\nu}/T_\gamma)^3 = 4/11$ for all neutrino types after the electron-positron annihilation. Thus, the electron neutrino asymmetry parameter can be estimated as~\cite{Serpico:2005bc,Lesgourgues:2006nd,Barger:2003rt}
\begin{align}\label{AsymSimp}
&\eta_{\nu_e}\simeq 0.249\xi_{\nu_e}\,,
\end{align}
where we only keep the leading-order term in $\xi_{\nu_e}$. According to the standard statistical physics, the equilibrium photon number density is given by
\begin{align}\label{PhotonDens}
&n_\gamma=\left(\frac{2\zeta(3)}{\pi^2}\right)T_\gamma^3\,.
\end{align}

By combining Eqs.~\eqref{AsymParam}, \eqref{AsymSimp} and \eqref{PhotonDens}, we can obtain the following electron neutrino asymmetry
\begin{eqnarray}\label{AsymDens}
\Delta n_{\nu_e} =\eta_{\nu_e} n_\gamma 
\simeq 0.061\xi_{\nu_e}T_\gamma^3
\end{eqnarray}
Predicted by the entropy conservation, the temperature of photons after their decoupling evolves as 
\begin{equation}\label{TempScaling}
 T_\gamma R = T_{\gamma\,0} R_0= T_{\gamma \,D} R_D\,,
\end{equation}
where $T_{\gamma \,D}$ ($T_{\gamma\,0}$) and $R_D$ ($R_0$) are the temperature and the scale factor at the time of recombination (at present). By defining the redshift $z$ as $R/R_0 \equiv 1/(1+z) $, the photon temperature at any redshift is given by
\begin{eqnarray}\label{TempZ}
 T_\gamma=T_{\gamma\,0}(1+z). 
\end{eqnarray}

By putting Eqs.~\eqref{AsymDens} and \eqref{TempZ} into the birefringence angle expression in Eq.~\eqref{AngleNu} and transforming the integration variable from $t$ to $z$ with
\begin{eqnarray}
	dt = \frac{d R}{HR} = - \frac{dz}{(1+z)H}\,,
\end{eqnarray}
we can obtain
\begin{eqnarray}\label{AngleNu1}
\Delta\alpha &=& 0.03 \beta \left( \frac{\xi_{\nu_e}T_{\gamma\,0}^3 }{M^2} \right) \int^{z_D}_0  \frac{(1+z)^2}{H(z)}dz  \nonumber\\
&\approx& 0.03 \beta \left(\frac{\xi_{\nu_e} T_{\gamma\,0}^3}{M^2 H_0}\right) \frac{2}{3} (1+z_D)^{3/2}\,,
\end{eqnarray}
where $z_D \simeq 1090$ denotes the redshift at the photon decoupling~\cite{ParticleDataGroup:2022pth} and we have approximated the cosmological evolution afterward as a flat and matter-dominant Universe with the Hubble parameter given by
\begin{eqnarray}\label{HubbleMatter}
&H(z)=H_0(1+z)^{{3}/{2}}\,,
\end{eqnarray}
in which the Hubble parameter $H_0$ and the CMB temperature $T_{\gamma\,0}$ at present~\cite{ParticleDataGroup:2022pth} are
\begin{eqnarray}\label{H0T0}
	H_0 = 100 h\, {\rm km \, s^{-1} Mpc^{-1}} \simeq 2.1332\times10^{-42}h \,\mbox{GeV}\,,\quad
 T_{\gamma\,0} \simeq 2.7255\,{\rm K}\,.
\end{eqnarray}
with $h\simeq0.674$ given by the {\it Planck} 2018 data.

\begin{figure}[ht]
\center
\includegraphics[width=0.7 \linewidth]{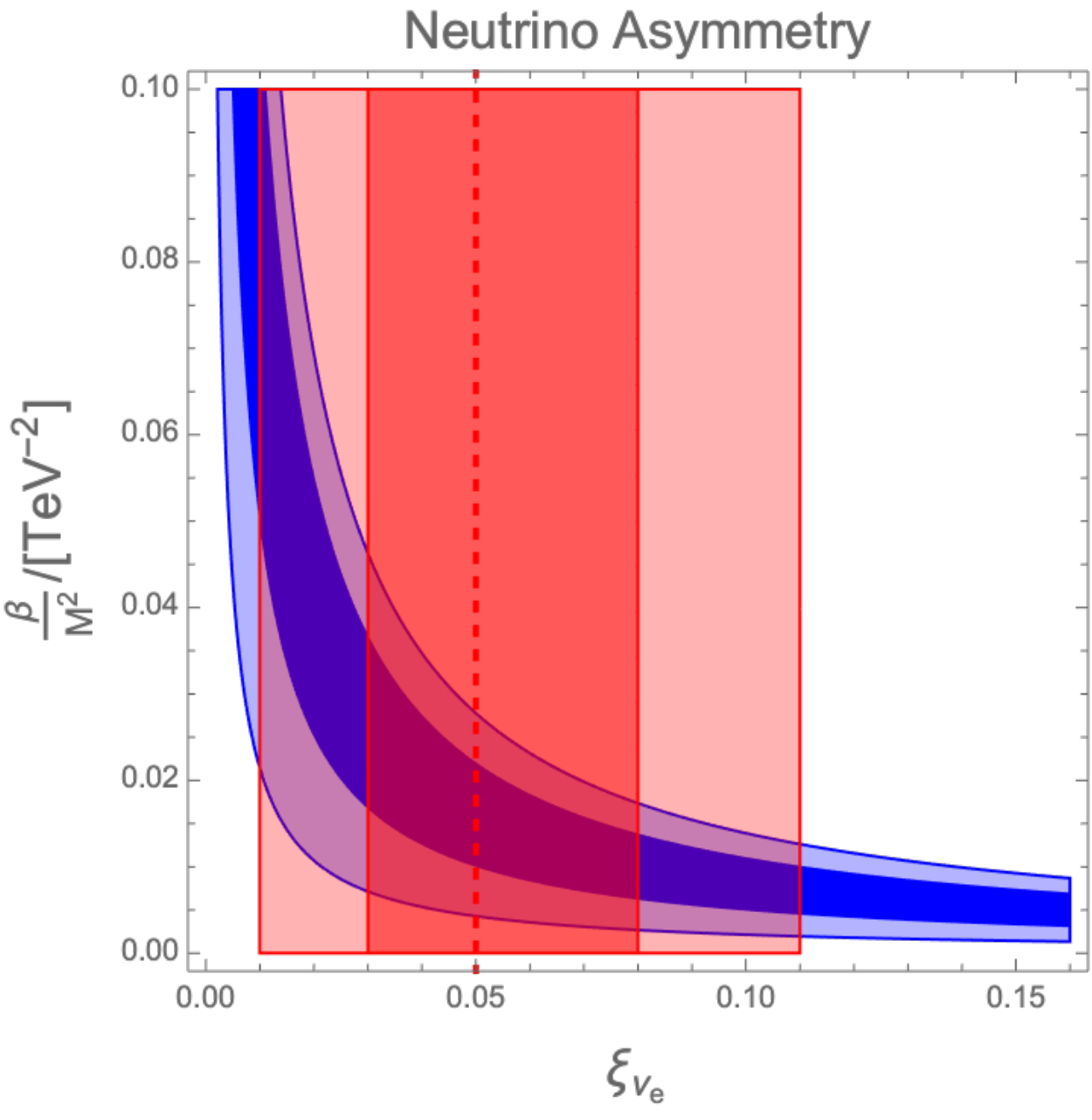}
\caption{{The parameter space in the $\xi_{\nu_e}$-$\beta/M^2$ plane, where the dark(light) blue region explains the CMB birefringence angle at 1$\sigma$(2$\sigma$) CL reported by Planck PR4, while the dark(light) red shaded area represents the $1\sigma$($2\sigma$) range of the electron neutrino asymmetry parameter $\xi_{\nu_e}$ indicated by the EMPRESS survey with the dotted line representing its central value.}}\label{FigNu}
\end{figure}

Note that the angle rotated by the CMB photon polarization plane reported by {\it Planck} PR4 is  $\Delta\alpha=0.30^{\circ}\pm0.11^{\circ} = (5.24 \pm 1.92)\times 10^{-3}$~rad at 68\% confidence level (CL)~\cite{Diego-Palazuelos:2022dsq}, which updates the analysis based on the {\it Planck} PR3 data in Ref.~\cite{Minami:2020odp}. Moreover, the electron neutrino degeneracy parameter $\xi_{\nu_e}$ reflecting the lepton asymmetry contained in $\nu_e$ is usually measured and constrained by the CMB and BBN observations. In particular, the latest measurement of the primordial helium abundance in the metal poor galaxies by the EMPRESS survey~\cite{Matsumoto:2022tlr} {has indicated that there could be a tension between the measured value of the primordial helium abundance and the standard BBN prediction, which might be explained by an exceptionally large nonzero electron neutrino asymmetry $\xi_{\nu_e}=0.05^{+0.03}_{-0.02}$~\cite{Matsumoto:2022tlr}. More recently, the joint analysis of BBN and CMB data in Refs.~\cite{Burns:2022hkq,Escudero:2022okz} have further strengthened the conclusion of a nonzero lepton asymmetry. If this anomaly persists, it possibly hints to new physics beyond the SM. In the present work, we would like to use the value of $\xi_{\nu_e} = 0.05^{+0.03}_{-0.02}$ given by the EMPRESS Collaboration for our following numerical analysis. As a result, the CMB polarization rotation angle  can be estimated as follows}
\begin{eqnarray}\label{33}
\Delta\alpha \simeq 5.24\times10^{-3}\beta\left(\frac{\xi_{\nu_e}}{0.05}\right)\left(\frac{7.9\,\mbox{TeV}}{M}\right)^2\,.
\end{eqnarray}
We also plot in Fig.~\ref{FigNu} the relevant parameter space in the $\xi_{\nu_e}$-$\beta/M^2$ plane, {in which the dark(light) blue region can explain the CMB birefringence angle at $1\sigma$($2\sigma$) CL reported by {\it Planck} PR4 while the dark(light) red shaded area is the $1\sigma$ ($2\sigma$) }interval of the electron neutrino degeneracy parameter indicated by the latest EMPRESS survey. 

{It is noted that the neutrino-photon interaction given in Eq.~\eqref{Lagrangian} could be already constrained by the existing experiments. One important constraint could be derived from the LEP search for the $Z$ boson rare decay $Z \to \nu\bar{\nu}\gamma\gamma$ with the $95\%$ CL upper limit on its branching ratio ${\cal B}(Z\to \nu\bar{\nu}\gamma\gamma) \lesssim 3.1\times 10^{-6}$~\cite{OPAL:1993ezs}. As a result, the upper bound on our $\nu\bar{\nu}\gamma\gamma$ coupling in Eq.~\eqref{Lagrangian} is given by 
\begin{eqnarray}
    \frac{\beta}{M^2}\leq  \frac{1.1 \times 10^3}{ \left({\rm TeV}\right)^{2}}\,  \quad \mbox{at 95\% CL,}
\end{eqnarray}
with its derivation presented in Appendix~\ref{SecLEP} where we also give a similar upper limit as in Ref.~\cite{Larios:2002gq} on the dimension-7 neutrino-diphoton interaction operators. Compared to the parameter space of phenomenological interest as shown in Fig.~\ref{FigNu}, this constraint is too weak to be displayed in the plot.  
Moreover, a stronger bound was given in Ref.~\cite{Gninenko:1998nn} by analyzing the Primakoff effect on the  $\nu_\mu Z \to \nu_s Z$ conversion process in the external Coulomb field of a nucleus $Z$ measured by the NOMAD Collaboration~\cite{NOMAD:1998pxi}. However, as seen from the experimental setup in the NOMAD neutrino detectors~\cite{NOMAD:1998pxi}, the targeted Primakoff process should have a muon neutrino in the initial state, whereas the explanation of the cosmic birefringence involves only $\nu_e$. Thus, the upper bounds yielded by NOMAD~\cite{Gninenko:1998nn} cannot be directly applicable to the present case. Finally, it has been proposed in Refs.~\cite{Sahin:2010zr,Sahin:2012zm,Senol:2012sn,Koksal:2023qch} that the sensitivity on the $\nu\bar{\nu}\gamma\gamma$ coupling could be significantly improved at the LHC as well as the planned CLIC and FCC-hh colliders. Nevertheless, to the best of our knowledge, the research for these collider channels only stays in theoretical studies, without any support of real experimental data.  }

\section{Asymmetric Dark Matter Current}\label{ADM}
{More and more astrophysical and cosmological evidence~\cite{ParticleDataGroup:2022pth,Planck:2018vyg} has shown the existence of the DM in our Universe(see {\it e.g.} Refs.~\cite{Bertone:2004pz,Bertone:2010zza,Bertone:2016nfn,deSwart:2017heh} for recent reviews and references therein)}, but its nature is still a great mystery. It is intriguing that the DM particle can be related to other beyond-SM physics, like the cosmic birefringence measured by {\it Planck}. In this section, we would like to interpret the birefringence angle in the CMB data as induced by the fermionic DM current $J^\chi_\mu = \bar{\chi} \gamma_\mu \chi$ through the effective interaction ${\cal L}_{\rm CS}$ in Eq.~\eqref{Lagrangian}. As shown in the Sec.~\ref{derivation}, it is the number density excess of the DM particles over anti-DM ones that sources the CMB polarization plane rotation in this setup. 
In the present work, we do not specify the origin of such DM asymmetries, and assume that all the DM density in the Universe is composed of the dark fermions without any corresponding anti-fermions, {\it i.e.}, $n_\chi = \Delta n_\chi = J^\chi_0$. Such a scenario is usually called the ADM model~\cite{Nussinov:1985xr,Kaplan:1991ah,Hooper:2004dc,Kaplan:2009ag,Petraki:2013wwa,Zurek:2013wia,Iminniyaz:2011yp,Graesser:2011wi,Ho:2022tbw}, {which, if the cosmological baryonic matter and DM densities originate from the same mechanism, could help to explain the observed cosmological density ratio of the visible and dark matters when the ADM mass is about 5 times of the proton/neutron mass}, {\it i.e.}, $M_\chi \approx 5$~GeV. 

\subsection{ADM Explanation of the Cosmic Birefringence}

At present, the DM abundance is usually parametrized by the following density parameter 
\begin{eqnarray}\label{OmegaX}
	\Omega_{\chi\,0} = \frac{\rho_{\chi\,0}}{\rho_{c\,0}} = \frac{ 8\pi G M_\chi n_{\chi\,0}}{3H_0^2} \,,
\end{eqnarray}
where $M_\chi$ denotes the ADM particle mass, while $H_0$, $\rho_c$, $\rho_{\chi\,0}$ and $n_{\chi\,0}$ are the present-day Hubble parameter, critical density, ADM mass density and its number density, respectively. Here we have used the relations $\rho_{\chi\,0} = M_\chi n_{\chi\,0}$ and $\rho_{c\,0} = 3H_0^2/(8\pi G)$ in the last equality with $G$ the Newton constant. By further considering the evolution of the ADM density with the cosmological expansion $n_\chi = (1+z)^3 n_{\chi\,0}$, we can obtain the birefringence angle induced by ADM as follows
\begin{eqnarray}\label{AngleADM}
	\Delta \alpha &=& \frac{1}{2} \frac{\beta}{M^2} \frac{\rho_{c\,0} \Omega_{\chi\,0}}{M_\chi} \int^{z_D}_0 (1+z)^3 \frac{dz}{H(1+z)} \nonumber\\
	&\approx& \frac{1}{2} \frac{\beta}{M^2} \frac{3H_0 \Omega_{\chi\,0}}{8\pi G M_\chi } \frac{2}{3} (1+z_D)^{3/2}\,,
\end{eqnarray}
where we also take into account the Hubble parameter evolution of $H=(1+z)^{3/2} H_0$ in the matter-dominated era. By taking the experimental values of various cosmological parameters~\cite{ParticleDataGroup:2022pth} into Eq.~\eqref{AngleADM}, the birefringence angle is given by
\begin{eqnarray}\label{AngleADM1}
	 \Delta\alpha =5.24\times10^{-3}\beta\left(\frac{1.77\, {\rm GeV}}{M}\right)^2\left(\frac{5\,{\rm GeV}}{M_{\chi}}\right)\,,
\end{eqnarray}  
where we have taken the benchmark ADM mass to be $M_\chi \approx 5$~GeV which could explain the cosmological ratio of the DM to the ordinary baryonic matter~\cite{Kaplan:2009ag}. 

In light of the expression of the photon polarization rotation angle $\Delta \alpha$ in Eq.~\eqref{AngleADM1}, {we would like to investigate existing experimental constraints on our ADM explanation of the CMB cosmic birefringence.} In fact, as shown in the following subsections, DM indirect and direct searches have already placed useful bounds on the relevant parameter space.

\subsection{Constraints From CMB Power Spectra}\label{SecCMB}
\begin{figure}[ht]
	\center
	\includegraphics[width=0.5 \linewidth]{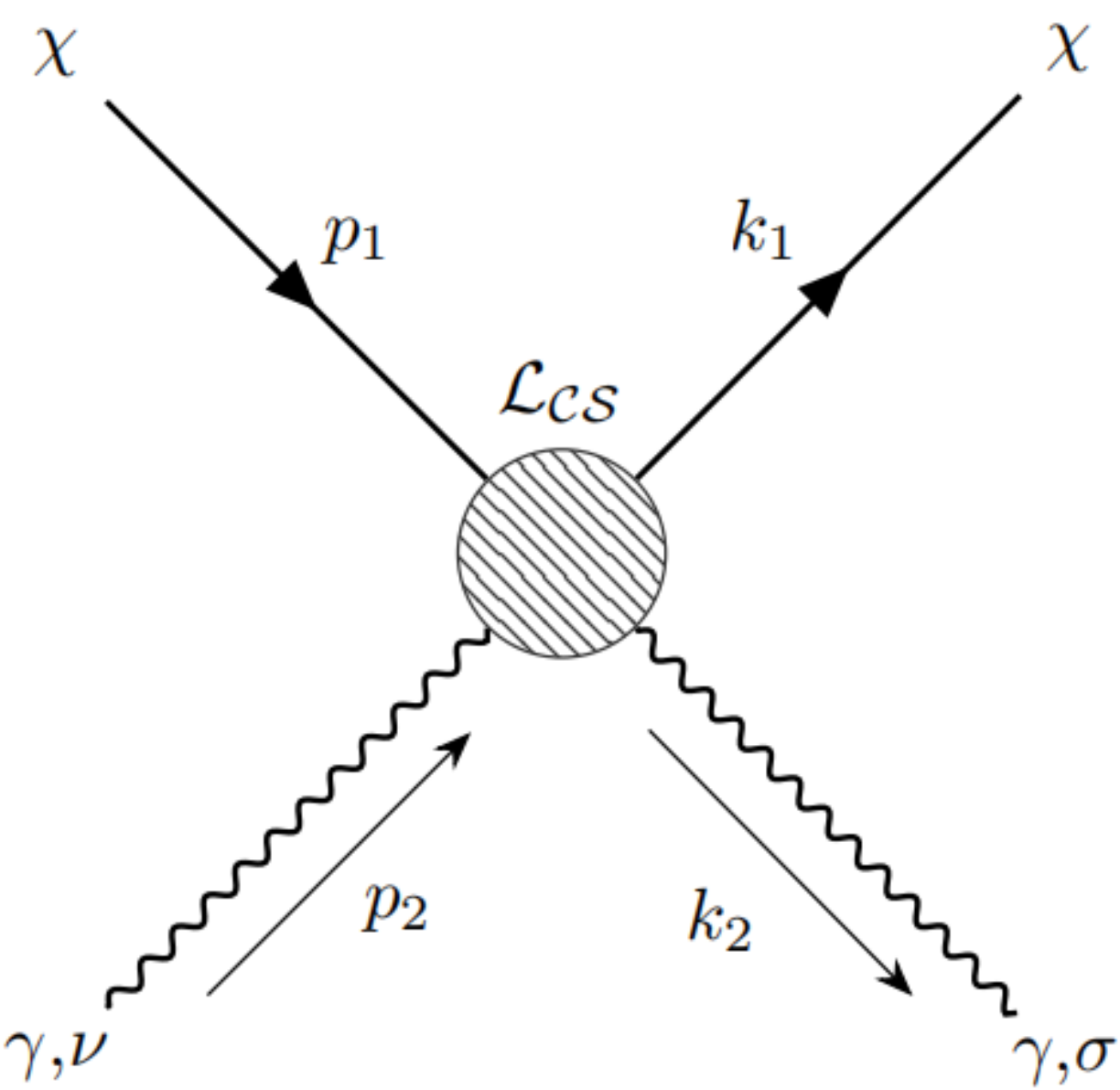}
	\caption{{The Feynman diagram for DM-photon scatterings.}}\label{FigGammaX}
\end{figure}

By identifying fermions in the current $J_\mu = \bar{\chi} \gamma_\mu \chi$ as ADM particles,  the effective CS interaction of ${\cal L}_{\rm CS}$ in Eq.~\eqref{Lagrangian} could generate ADM-photon elastic scatterings as shown in Fig.~\ref{FigGammaX}, which would leave imprints on the CMB angular spectrum and the large scale structure~\cite{Boehm:2001hm,Wilkinson:2013kia,Stadler:2018jin,Becker:2020hzj,Zhou:2022ygn}. In particular, due to the collisional damping caused by the ADM-photon interaction,  the obtained matter power spectrum would show significant suppression at small scales together with a series of damped oscillations. Moreover, such a scattering between ADM and photons would also manifest itself in the CMB power spectra as modifications of relative magnitudes and shifts of positions of the acoustic peaks. Therefore, we can use the CMB angular power spectra of temperature and polarizations to constrain our ADM model. 

Note that the ADM-photon interaction in Eq.~\eqref{Lagrangian} gives rise to the following amplitude
\begin{align}\label{AmpXG}
	i\mathscr{M} = \frac{1}{2}\frac{\beta}{M^2}\bar{u}_\chi(k_1)\gamma_{\mu}u_\chi(p_1) \epsilon^{\mu\nu\rho\sigma}(k_2+p_2)_{\rho} \epsilon_\nu(p_2)\epsilon^{\ast}_\sigma(k_2)\,,
\end{align}
which  leads to the ADM-$\gamma$ scattering cross section, given by
\begin{align}\label{XecXG}
	\sigma_{\chi\gamma} \approx \frac{\beta^2 p_{1\,\mathrm{cm}}^2}{8\pi M^4}\,,
\end{align}
where $p_{1\,\mathrm{cm}} = |\mathbf{p}_{1\,\mathrm{cm}}|$ stands for the incoming photon momentum in the center-of-mass (cm) frame. By assuming that the ADM has already become non-relativistic around the photon decoupling, the relation $M_\chi \gg p_{1\,\mathrm{cm}} \sim T_\gamma$ holds so that we only keep the leading-order term in the expansion with respect to the small ratio of $p_{1\,\mathrm{cm}}/M_\chi$ in Eq.~\eqref{XecXG}. 

It is shown in Refs.~\cite{Kolb:1990vq,Gondolo:1990dk} that the quantity controlling the cosmological evolution of ADM and photons in the Boltzmann equations is the following thermally averaged ADM-photon cross section
\begin{align}\label{DefXSv}
	\langle \sigma v_{\rm M\o l} \rangle_{\chi \gamma} = \frac{\int \sigma_{\chi\gamma} v_{\rm M\o l} d n_\gamma d n_\chi}{\int d n_\gamma dn_\chi}\,,
\end{align}
where $v_{\rm M{\o}l}$ is the M{\o}ller velocity~\cite{Gondolo:1990dk} and the differential density $d n_i$ is defined by
\begin{eqnarray}
	d n_{i} = g_i\frac{d^3 p_i}{(2\pi)^3} f_i (p_i)\,,
\end{eqnarray} 
in which $g_i$ is the independent degrees of freedom of the particle $i$ and $f_i (p_i)$ is the associated distribution. Here, the distributions for photons and ADM particles are defined in the cosmic comoving frame. Since photons are always kept in the thermal equilibrium state with temperature $T_\gamma$, so that they should obey the Bose-Einstein distribution
\begin{eqnarray}\label{GammaDist}
	 f_\gamma (p) = \frac{1}{e^{p/T_\gamma}-1}\,, 
\end{eqnarray}
where we have taken the Boltzmann constant to be $k_{\rm B} = 1$. For the ADM, we do not need the explicit form of its distribution function $f_\chi (p)$ here. As argued in Ref.~\cite{Gondolo:1990dk}, due to the  relation
\begin{eqnarray}\label{LabDef}
	\langle \sigma v_{\rm M{\o}l} \rangle = \langle \sigma v_{\rm lab} \rangle^{\rm lab}\,,
\end{eqnarray}
it is more convenient to compute the thermally averaged cross section in the lab frame, in which the ADM particle in the scattering is initially at rest. In Eq.~\eqref{LabDef}, $v_{\rm lab}$ refers to the relative velocity and the superscript ``lab'' on the bracket denotes the thermal average computed in the lab frame. Also, at the leading order in the small momentum expansion, the ADM-photon scattering cross section in the lab frame takes the same form as in Eq.~\eqref{XecXG} except for the photon momentum $p_{1\,{\rm cm}}$ replaced by the counterpart $p_{1\,{\rm lab}}$. Therefore, by taking Eqs.~\eqref{GammaDist} and \eqref{XecXG} into Eq.~\eqref{LabDef}, the thermally averaged ADM-photon cross section is given by
\begin{eqnarray}\label{XGXecT}
	\langle \sigma v_{\rm M{\o}l} \rangle_{\chi\gamma} \simeq \frac{3\zeta(5)}{2\pi \zeta(3)} \frac{\beta^2 T_\gamma^2}{M^4} = 0.412 \left( \frac{\beta^2 T_\gamma^2}{M^4}\right)\,,
\end{eqnarray}
where we have only kept the dominant term when $T_\gamma \ll M_\chi$. In the derivation of Eq.~\eqref{XGXecT}, we have factored out and cancelled the ADM density $n_{\chi}$ between the numerator and denominator in Eq.~\eqref{DefXSv} since $\sigma v_{\rm lab}$ does not depend on the ADM momentum at all. 

{For the given ADM-photon scattering} with the cross section quadratically proportional to the photon temperature $T_\gamma$, the best constraint is given in Ref.~\cite{Wilkinson:2013kia} as follows
\begin{eqnarray}\label{CMBBound}
	\langle\sigma v_{\rm M\o l} \rangle_{\chi\gamma} (T_{\gamma 0}) \lesssim 6\times 10^{-40} \left(\frac{M_\chi}{\rm GeV}\right) {\rm cm}^2\,, \quad \mbox{at 68\% C.L.}\,,
\end{eqnarray}
{where  $T_{\gamma0} \simeq 2.7255$~K~\cite{ParticleDataGroup:2022pth} is the present-day CMB temperature.} By comparing Eqs.~\eqref{XGXecT} and \eqref{CMBBound}, we can express the CMB constraint in terms of our model parameters as follows
\begin{eqnarray}\label{CMBBound1}
	\frac{\beta}{M^2} \lesssim 8.24 \times 10^6\, \mbox{GeV}^{-2} \left(\frac{M_\chi}{\rm GeV}\right)^{1/2}\,.
\end{eqnarray}
As a result, given parameters in Eq.~\eqref{AngleADM1} required to explain the observed cosmic birefringence, the limit in Eq.~\eqref{CMBBound1} is too weak to place any useful constraint on the model parameters, especially when the ADM particle mass is taken to be $M_\chi \lesssim 10$~GeV.  

Finally, we would like to mention that  the ADM-photon coupling upper bound presented in Eq.~\eqref{CMBBound} was derived in Ref.~\cite{Wilkinson:2013kia} by using the {\it Planck} 2013 data on the CMB $TT$ and $EE$ auto power spectra, {which are somewhat out of date.} In particular, the {\it Planck} Collaboration has released their data on the CMB angular power spectra of temperature, polarization and lensing in 2015 and 2018. Moreover, as shown in Refs.~\cite{Zhou:2022ygn,Becker:2020hzj}, the inclusion of the data from BAO and weak lensing experiments can further strengthen the constraining power. Therefore, we expect that the CMB constraint on the ADM-photon interaction in Eq.~\eqref{CMBBound1} can be further improved by updating the CMB data along with the BAO and weak lensing data. Unfortunately, such a goal has only been achieved in Refs.~\cite{Stadler:2018jin,Becker:2020hzj,Zhou:2022ygn} for the case with a constant DM-photon scattering cross section. For the present ADM model with the photon scattering cross section proportional to $T_\gamma^2$, there is not any new progress after the study in Ref.~\cite{Wilkinson:2013kia}, which provides the best experimental limit up to now.  


\subsection{Constraint From DM Direct Detections}
\begin{figure}[ht]
	\center
	\includegraphics[width=0.5\linewidth]{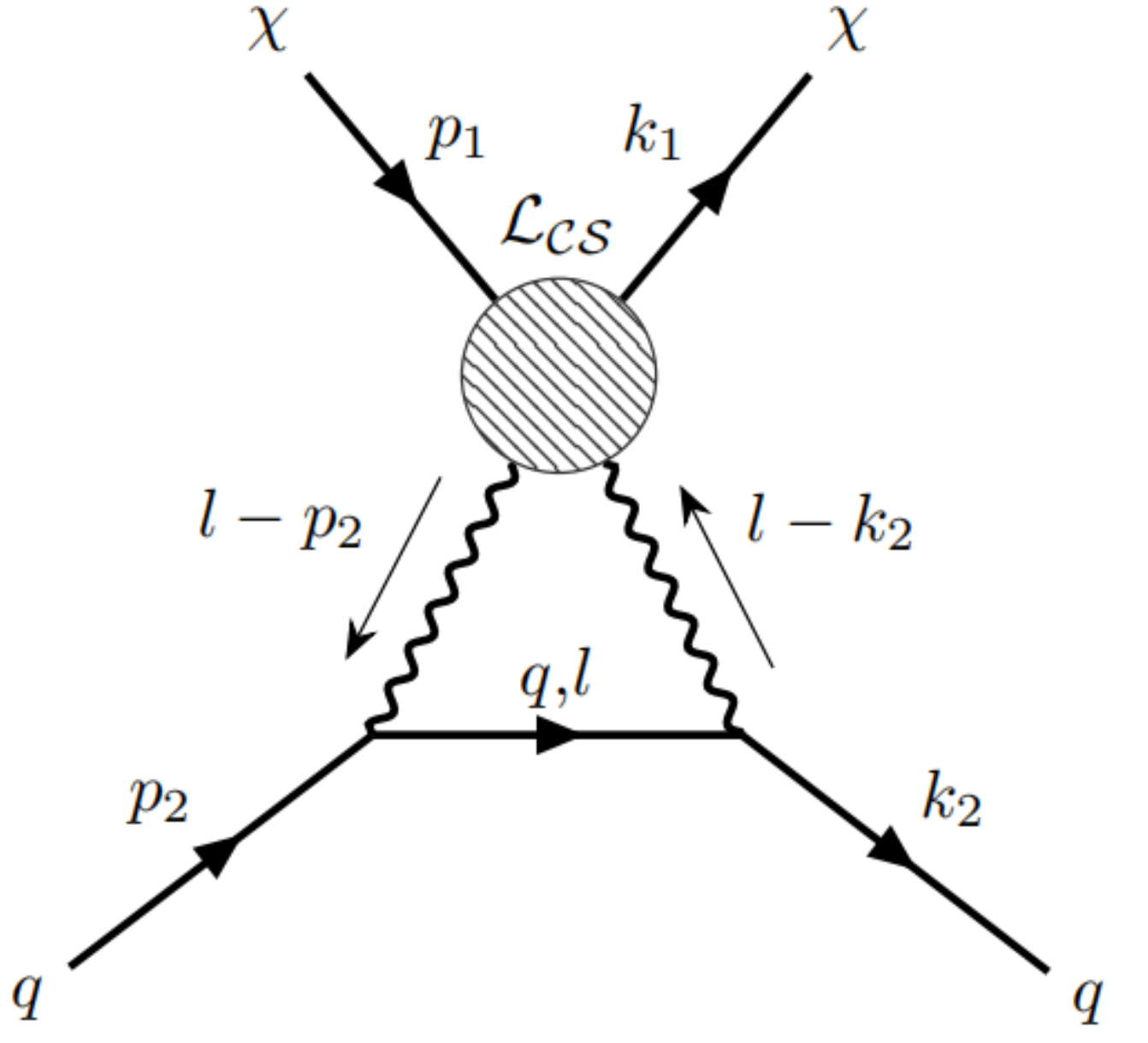}
	\caption{The Feynman diagram for ADM direct detections.}\label{FigDMDD}
\end{figure}

The ADM-photon interaction in Eq.~\eqref{Lagrangian} can also induce the effective couplings between the ADM particle $\chi$ and SM quarks at the one-loop level as illustrated in Fig.~\ref{FigDMDD},
which can be further probed by the DM DD experiments. Note that the loop integral for the Feynman diagram of Fig.~\ref{FigDMDD} is logarithmically divergent due to the insertion of the nonrenormalizable ADM-photon effective operator ${\cal L}_{\rm CS}$. Therefore, it is expected that the ADM-quark scattering is dominated by the logarithmically divergent term, which can be expressed by the following effective ADM-quark interaction
\begin{eqnarray}\label{OXq}
	{\cal L}_{\chi q} = -\sum_q \frac{1}{m^{2}_{V_q}} \bar{\chi}\gamma_\mu \chi \bar{q}\gamma^\mu \gamma^5 q\,,
\end{eqnarray} 
where
\begin{eqnarray}\label{WCXq}
	\frac{1}{m^2_{V_q}} = \frac{3\alpha}{8\pi} \frac{\beta}{M^2}  Q_q^2 \ln\frac{\Lambda^2}{m_q^2}\,,
\end{eqnarray}
with $m_q$ and $Q_q$ denoting the mass and charge of the quark flavor $q$, respectively, while other contributions are suppressed by small scales such as the momentum transfer or external particle momenta. The factor of $\ln \Lambda^2/m_q^2$ comes from the logarithmic divergence with the UV cutoff scale identified as $\Lambda$, which can be equal to $M$ or not depending on model assumptions. Also, we have followed the convention in Ref.~\cite{DEramo:2016gos} to parametrize the ADM-quark couplings to be $1/m^2_{V_q}$, as if there is a heavy vector particle $V_q$ of mass $m_{V_q}$ mediating the interaction between the flavor $q$ and $\chi$.  

In order to connect the effective interactions in Eq.~\eqref{OXq} with the observables in the DM direct searches, one can match ${\cal L}_{\chi q}$ to the nucleon-level non-relativistic (NR) operators ${\cal O}_7^{\rm NR}$ and ${\cal O}_9^{\rm NR}$~\cite{Fitzpatrick:2012ix,Fitzpatrick:2012ib,Anand:2013yka,DEramo:2016gos}, both of which lead to velocity and momentum suppressed spin-dependent ADM-quark interactions. Hence, we expect na\"ively that the DM DD constraints imposed on operators in Eq.~\eqref{OXq} would be extremely weak. However, as pointed out in Refs.~\cite{DEramo:2016gos,Bishara:2018vix,Baumgart:2022vwr}, 
the renormalization group (RG) running would cause the mixing among dimension-six DM-quark effective operators and, in particular, generate the couplings of the DM vector current to the quark vector current, such as $\bar{\chi} \gamma_\mu \chi \bar{q} \gamma^\mu q$, which would further induce the spin-independent ADM-quark scatterings without any velocity or momentum suppression. The latter scattering channel would be even enhanced by the coherence of the large number of nucleons in the target heavy nucleus. As a result, such an operator mixing effect would significantly strengthen the DD constraints on our ADM model.     

Currently, the best DM DD constraint comes from the LUX-ZEPLIN (LZ) Collaboration~\cite{LZ:2022ufs}, which has excluded  the DM-nucleon elastic spin-independent cross section larger than $6.5\times 10^{-48}\,{\rm cm}^2$ when $M_\chi = 30$~GeV at $90$\% CL. The detailed computation of the DD exclusion limit on our ADM model parameters, as well as the relevant RG running and mixings of the dimension-six ADM-quark interaction operators, is well beyond the scope of the present work. Instead, we would like to apply the estimated LZ constraint in Ref.~\cite{DEramo:2016gos},  which has shown lower bounds on the heavy mediator mass $m_V$ as a function of DM mass $M_\chi$ in the lower left panel of Fig.~1 {of Ref.~\cite{DEramo:2016gos}}~\footnote{We have used {\sf WEBPLOTDIGITIZER}~\cite{Rohatgi2022} to obtain the reported LZ limits on the heavy mediator mass $m_V$ by digitizing the lower-left plot of Fig.~1 in Ref.~\cite{DEramo:2016gos}.}. 
However, we should take care of several differences in model assumptions in Fig.~1 of Ref.~\cite{DEramo:2016gos} from our present ADM setup. {Firstly, the DM DD limit on the DM-quark operators in Sec.~4.1} of Ref.~\cite{DEramo:2016gos} was placed on the case in which all quark flavors share a universal coupling, {which means that the effective heavy mediator masses $m_{V_q}$ in Eq.~\eqref{OXq} take the same value $m_{V_q} = m_V$}.  In contrast, our ADM-quark effective interaction of each quark flavor has its respective coupling $1/m^2_{V_q}$ which is caused by the quark electric charge $Q_q$ and mass $m_q$ as expressed in Eq.~\eqref{WCXq}. Thus, in general, we cannot apply the result in Fig.~1 of Ref.~\cite{DEramo:2016gos} directly to our ADM model. Nevertheless, here we are only interested in the order estimation of the DM DD bounds, while the variations in quark electric charges and masses only give rise to  ${\cal O}(1)$ corrections to the DD constraints, which can be illustrated by the similar result for the third-generation quark-DM coupling presented in the lower-left plot in Fig.~5 of Ref.~\cite{DEramo:2016gos}. As a result, we can simplify our analysis of the effective operators of Eq.~\eqref{OXq} by approximating in Eq.~\eqref{WCXq} all quark charges to be $Q_q= 2/3$ and the logarithmic factor to be $\ln \Lambda^2/m_q^2 \sim 5$. A further complication comes from the fact that the lower bounds on $m_V$ in Ref.~\cite{DEramo:2016gos} are estimated by assuming the experimental exposure to be $\omega = 5600$~ton\,yr~\cite{LZ:2015kxe}, while the LZ first search results were presented based on the 60 live-day data using the 5.5~ton fiducial-mass xenon~\cite{LZ:2022ufs} from which the exposure can be derived to be $\omega_0 = 0.93$~ton\,yr. One can account for this issue by simply rescaling the original lower $m_V^{\rm LZ}$ bounds as $m_V^{{\rm LZ}\,0} = m_V^{\rm LZ} (\omega_0/\omega)^{1/4}$ since the reduction of the LZ exposure would {loosen the DD upper limits} on the DM-nucleon cross section by a factor of $\omega/\omega_0$. {Note that here we have not considered the impact of the irreducible coherent neutrino background, the so-called neutrino floor, on the LZ sensitivity curve and the obtained DD constraints. On the one hand, the DM DD limits in Ref.~\cite{DEramo:2016gos} was given without taking this issue into account either. On the other hand, as shown in Ref.~\cite{LZ:2022ufs}, the coherent elastic neutrino-nucleus scattering rate contributed little to the expected nucleus scattering under the current experimental exposure and event selection thresholds imposed by the LZ Collaboration.} Finally, the DM DD bounds have been given for the DM mass above 10~GeV in Fig.~1 of Ref.~\cite{DEramo:2016gos} while the real LZ or other xenon-based DD experiments can extend their sensitivity to even lower DM mass range. Here we approximate the DM DD constraints for $M_\chi$ below 10~GeV by {extrapolating} the exclusion limits on $m_V^{{\rm LZ}\,0}$ to lower DM mass regions. {Concretely, the LZ projected sensitivity curve  in the low DM mass region around $M_\chi \approx 10$~GeV in the lower-left panel in Fig.~1 of Ref.~\cite{DEramo:2016gos} can be fitted by a linear function in the $\ln M_\chi$-$\ln m_V$ plane. The LZ constraint at lower $M_\chi$ can be yielded by extending this linear function into the DM mass region below 10~GeV.} With the above treatments and approximations, we can obtain the estimated DM DD constraints on the ADM models, which will be imposed on the parameter space in the following numerical analysis.




\subsection{Benchmark Scenarios}
Rather than scanning over the whole parameter space for the ADM model, we would like to numerically explore two specific benchmark ADM cases with the ADM mass of $M_\chi = 5$\,GeV and 5\,keV. The former is motivated by the possible explanation of the measured DM-baryon mass density ratio in our Universe, whereas the latter would provide us a potential warm DM candidate~\cite{Bode:2000gq} which could solve so-called ``Missing Satellites Problem''~\cite{Klypin:1999uc}. 

\begin{figure}[ht]
	\center
	\includegraphics[width=0.6\linewidth]{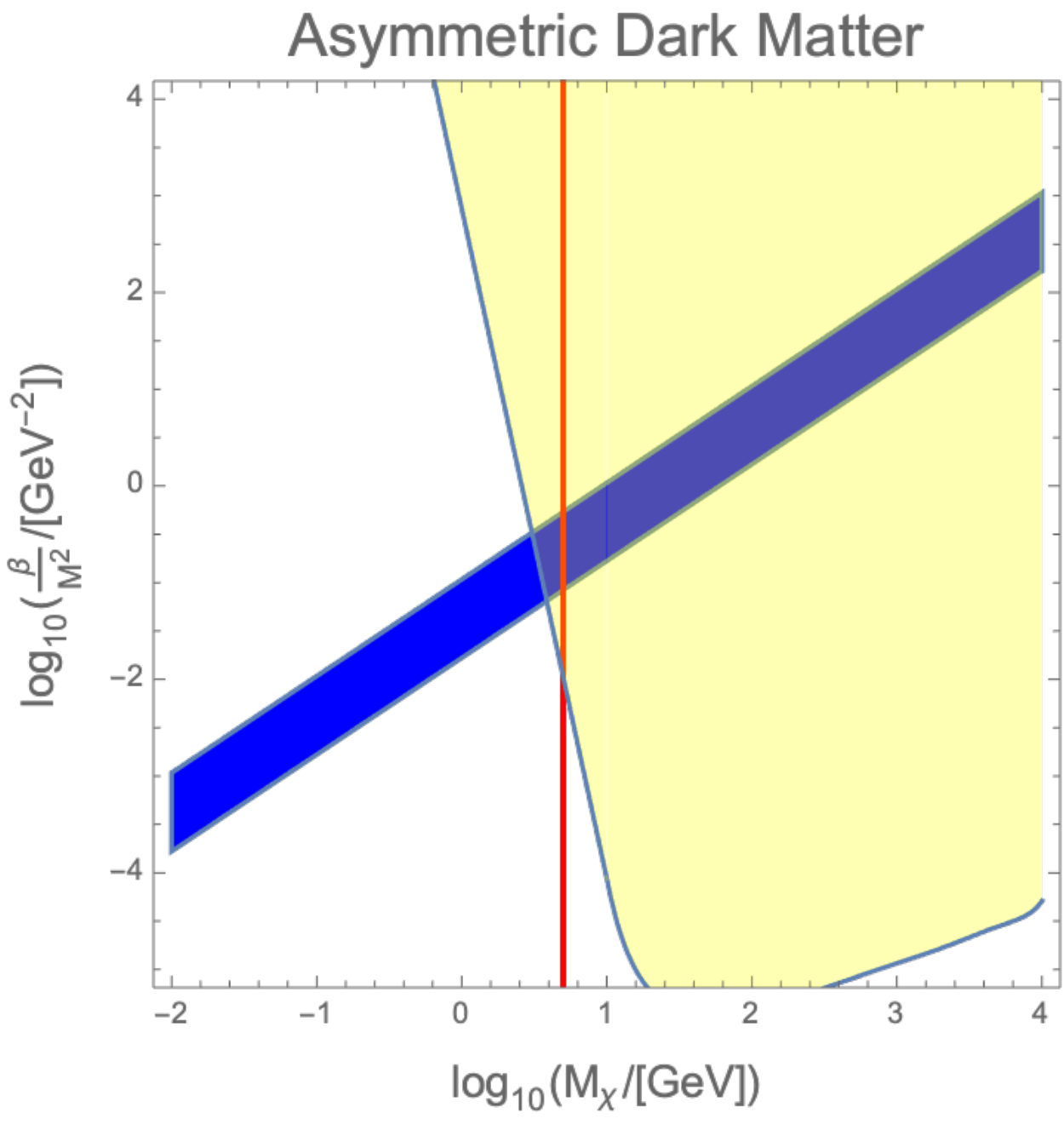}
	\caption{The parameter space in the $\log_{10}(M_\chi/{\rm GeV})$- $\log_{10} (\beta/M^2 /({\rm GeV})^{-2})$ plane. The solid blue band illustrates the parameter region favored by the cosmic birefringence signal observed by Planck, while the yellow area has been excluded by LZ experiment. The vertical red line denotes the ADM mass to be $M_\chi = 5$~GeV, which is preferred by the observed DM-baryon mass ratio in our Universe. }\label{PlotADM}
\end{figure}

In Fig.~\ref{PlotADM}, {we illustrate the numerical results for the ADM mass $M_\chi = 5$\,GeV}. The solid blue band represents the parameter region explaining the cosmic birefringence observed by {\it Planck}, while the yellow shaded area has been excluded by the LZ DD constraints. The red line denote the ADM mass of $M_\chi = 5$\,GeV. It is shown that, for a GeV-scale ADM particle, the favored region by the cosmic birefringence has totally ruled out by current DM DD experiments.  {Here we would like to mention that our present estimate of the DM DD limit is rather crude. For example, we have not appropriately addressed the neutrino floor issue. Nevertheless, we believe that the LZ limit obtained by our approach is conservative, and a more careful treatment of LZ data should even strengthen the constraint and thus reinforce our conclusion. Moreover,} as mentioned in subsection \ref{SecCMB}, the CMB constraint in Eq.~\eqref{CMBBound1} is so weak that we do not show it here.

On the other hand, when we lower the ADM mass to the warm DM region, {\it e.g.}, $M_\chi = 5$\,keV, the interpretation of the cosmic birefringence signal in the {\it Planck} PR4 data requires the effective coupling of ${\cal L}_{\rm CS}$ to be around $\beta /M^2 = 1/(1.77~{\rm TeV})^{2}$ based on Eq.~\eqref{AngleADM1}. Note that the stringent DM DD constraint from the LZ experiment cannot be applied to the present case since the DD technique loses its sensitivity in such a low DM mass region. In addition, the detection of CMB spectral distortions from the FIRAS data~\cite{Fixsen:1996nj} could provide an even stronger upper bound on the ADM-photon scattering cross section~\cite{Ali-Haimoud:2015pwa,Bianchini:2022dqh}
\begin{eqnarray}
	\sigma_{\chi\gamma}^0  \lesssim 2\times 10^{-37} {\rm cm}^2 \left(\frac{M_\chi}{\rm MeV}\right) \left( \frac{E_0}{E_\gamma}\right)^2\,,
\end{eqnarray}
for $M_\chi$ in the range of about 1~keV to 100~keV, where $\sigma_{\chi \gamma}^0$ denotes the cross section at the photon energy $E_0 = 1$~keV. Using Eq.~\eqref{XecXG} with $E_\gamma = p_{1\,{\rm cm}}$, the above constraints can be expressed by our ADM model parameters as follows
\begin{eqnarray}
	\frac{\beta}{M^2} \lesssim \left( \frac{M_\chi}{5\,{\rm keV}}\right)^{1/2} \left(\frac{1}{1.2\times 10^{-4}\, {\rm TeV}}\right)^2\,,
\end{eqnarray}
Obviously, such a constraint is still too weak compared with the parameters obtained by the measured cosmic birefringence angle. Finally, the ADM mass of $M_\chi = 5$~keV is well above the lowest DM mass bound $M_\chi \gtrsim 1$~keV derived from the phase space density considerations in Ref.~\cite{Boyarsky:2008ju}. Note that there are other stringent constraints from observations of the Lyman-$\alpha$ forest~\cite{Boyarsky:2008xj} and the matter power spectrum~\cite{Boehm:2014vja,Escudero:2018thh,Maamari:2020aqz}, which can also limit the DM mass and the ADM-photon interactin in Eq.~\eqref{Lagrangian}. However, such constraints are rather indirect, and contain many uncertainties from non-linear matter evolutions. Therefore, in the present work, we do not consider their impacts on our model parameters. 

\section{Conclusions}\label{conclusion}
Motivated by the recent measurement of the potentially nonzero cosmic birefringence angle from the {\it Planck} PR4 data, we consider its possible origin from the  CS-like coupling ${\cal L}_{\rm CS}$ of photons with some fermionic currents, which was previously studied in Ref.~\cite{Geng:2007va}. We have first revisited the derivation of the general formulae for the cosmic birefringence angle $\Delta \alpha$ from this photon-fermion effective operator, correcting a mistake in the corresponding expression in Ref.~\cite{Geng:2007va}. We have then identified the fermion in the current as the left-handed electron neutrino and the DM, and discuss their respective phenomenology. For the electron neutrino case, with the updated value of the degeneracy parameter $\xi_{\nu_e}$ from the EMPRESS survey~\cite{Matsumoto:2022tlr}, the explanation of the birefringence requires $\beta/M^2 \simeq 1/(7.9~{\rm TeV})^2$. On the other hand, if the current is assumed to be composed of fermionic DM particles, the birefringence angle should be proportional to the abundance difference between DM and anti-DM particles. For simplicity, we have further assumed that only DM particles are left in the present-day universe. In other words, we need to consider the ADM model. We have also explored the experimental constraints from the CMB power spectra and the DM direct searches. As a result, the CMB limit on the ADM model is too weak to be relevant in constraining the model parameter space, while the DM DD bounds from the LZ Collaboration have totally excluded the parameter region for $M_\chi \sim 5$~GeV, which is the natural ADM mass value so as to explain the observed cosmic DM-baryon mass density ratio. In contrast, for the warm DM case with $M_\chi \sim 5$~keV, the measured value of $\Delta \alpha$ from {\it Planck} 2018 data has restricted the effective coupling to be around $\beta/M^2 \simeq 1/(1.77\,{\rm TeV})^{2}$, which is shown to satisfy all the present experimental constraints.

\appendix
\section{Estimations of LEP Constraints on Effective Neutrino-Photon Interactions}\label{SecLEP}
{In this appendix, we shall present a simple estimate of the LEP constraints for the effective $\nu\bar{\nu}\gamma\gamma$ couplings based on the OPAL data on the $Z$-boson rare decay channel $Z\to \nu\bar{\nu}\gamma\gamma$~\cite{OPAL:1993ezs}. The corresponding $95\%$ CL upper limit on the branching ratio of this process is given by ${\cal B}(Z\to \nu\bar{\nu}\gamma\gamma) \leq 3.1\times 10^{-6}$. In order to show the validity of our method, we shall begin our discussion by considering the following dimension-7 operator~\cite{Larios:2002gq}
\begin{eqnarray}\label{NuFF}
	\mathcal{L}_{\nu_i\bar{\nu}_j\gamma\gamma} = \frac{1}{4\Lambda^3}\bar{\nu_i}(\alpha^{ij}_LP_L+\alpha^{ij}_RP_R)\nu_j\tilde{F}^{\mu\nu}F_{\mu\nu}\,,
\end{eqnarray}
where $\alpha^{ij}_{R,L}$ are dimensionless coupling constants with $i,\,j$ denoting the flavor indices, while $\Lambda$ represents the new physics cutoff scale. Such a $\nu\bar{\nu}\gamma\gamma$ effective interaction could induce the rare decay $Z\to \nu\bar{\nu}\gamma\gamma$ with Feynman diagrams shown in Fig.~\ref{FigZD}.
\begin{figure}[ht]
	\center
	\subfigure[]{
		\includegraphics[width=0.4\linewidth]{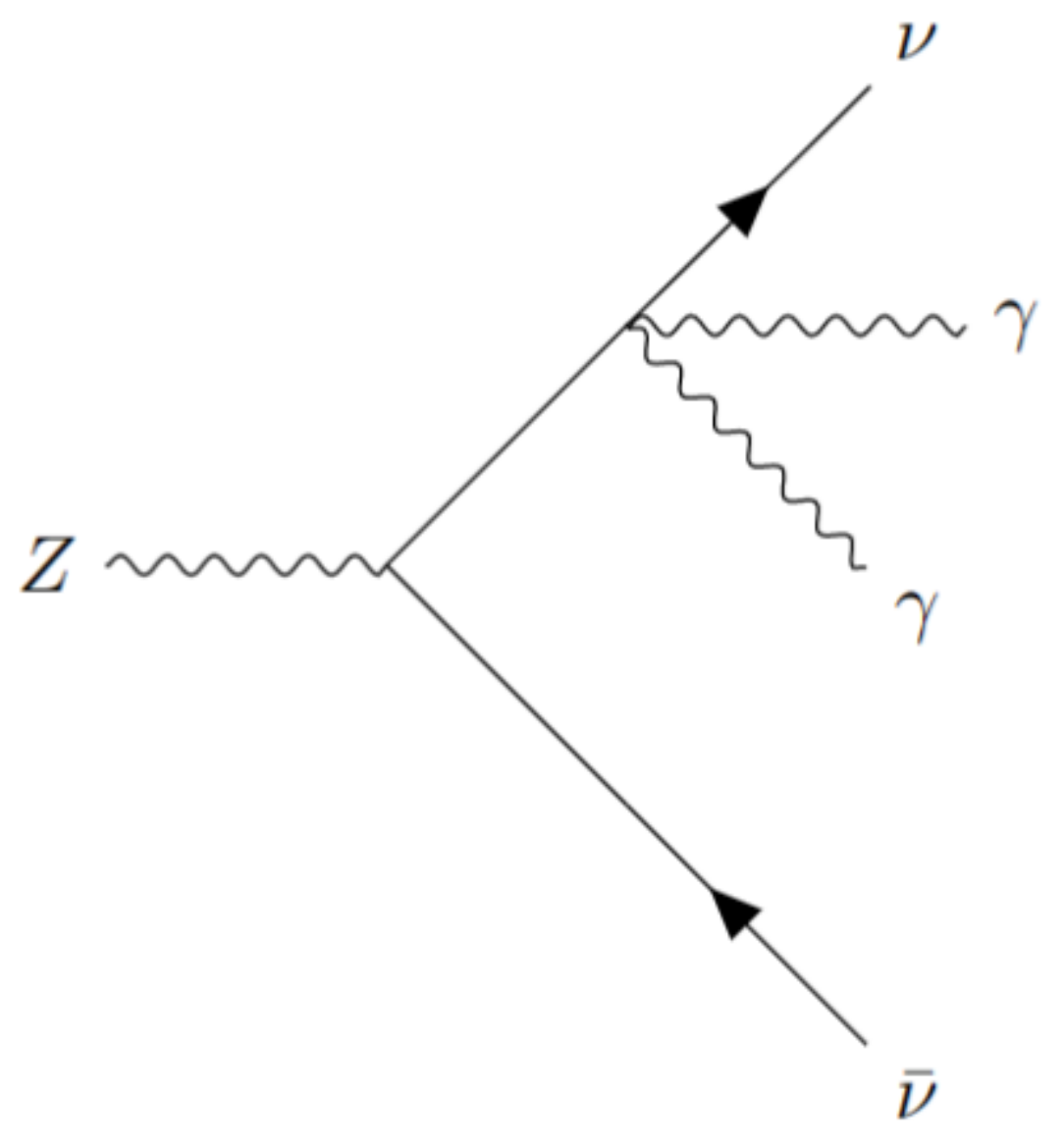}}
	\subfigure[]{
		\includegraphics[width=0.4\linewidth]{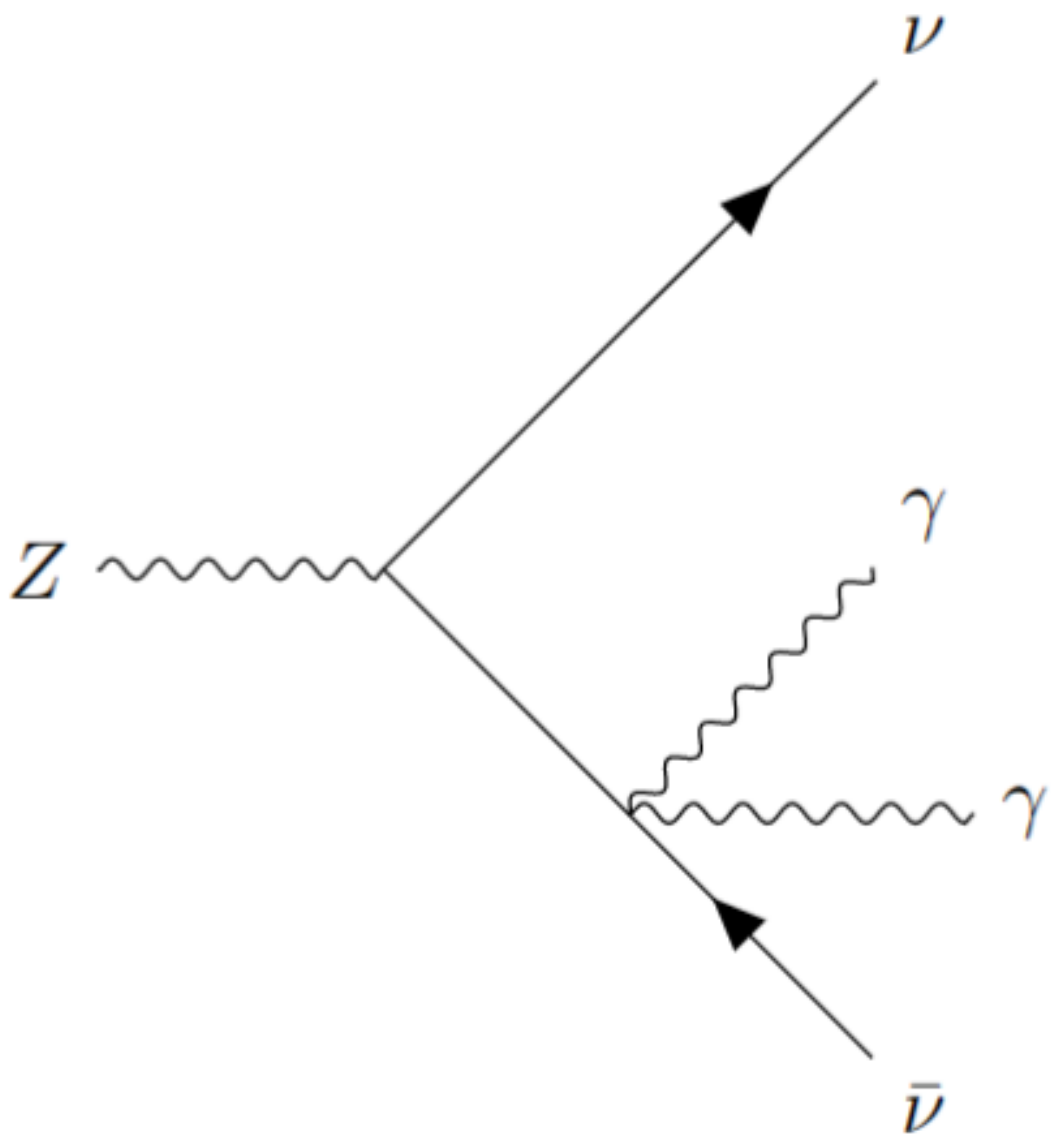}}
	\caption{{Feynman diagrams for  $Z\rightarrow\nu\bar{\nu}\gamma\gamma$ induced by effective $\nu\bar{\nu}\gamma\gamma$ interactions.}}\label{FigZD}
\end{figure}\\
As a result, we can estimate the branching ratio of this process by comparing it with that of the $Z\to \nu\bar{\nu}$ decay mode as follows
\begin{eqnarray}\label{BrDim7}
	{\cal B} \left(Z\to \nu\bar{\nu}\gamma\gamma\right) \approx {\cal B} \left(Z\to \nu\bar{\nu}\right)\times 4 \times\frac{1}{\left(4\pi\right)^4}\times\sum\limits_{i,j}\left(\left|\alpha^{ij}_R\right|^2+\left|\alpha^{ij}_L\right|^2\right)\left(\frac{E_{\nu,\gamma}}{\Lambda}\right)^6\,,
\end{eqnarray} 
where ${\cal B} (Z\to \nu\bar{\nu}) \simeq 0.067$ is the branching ratio of the $Z$-boson decay into a pair of neutrinos of a fixed flavor~\cite{ParticleDataGroup:2022pth}. In Eq.~\eqref{BrDim7}, the factor of $1/(4\pi)^4$ accounts for the phase space suppression for the two additional photons in the final state~\cite{Han:2005mu}, while the number 4 is derived from the fact that there are two Feynman diagrams with the effective operator in Eq.~\eqref{NuFF} inserting into either the outgoing neutrino or anti-neutrino line. In order to balance the mass dimension to make right-hand side of Eq.~\eqref{NuFF} dimensionless, we have introduced a energy scale $E_{\nu,\gamma}$, which is taken to be the averaged energy of the external particles $E_{\nu,\gamma} = m_Z/4$ with the $Z$-boson mass $m_Z= 91.1876$~GeV. In comparison with the 95\% CL upper bound given by the LEP measurement of this rare decay mode, the constraint on effective couplings in Eq.~\eqref{NuFF} can be yielded as follows
\begin{eqnarray}
	\left[ \frac{\rm 1~GeV}{\Lambda}\right]^6 \sum\limits_{i,j} \left(|\alpha^{ij}_L|^2 + |\alpha^{ij}_R|^2\right) \lesssim 2.1 \times 10^{-9} \quad \mbox{ at 95\% CL} \,.
\end{eqnarray}
Note that this result is in good agreement with the more complete numerical analysis given in Ref.~\cite{Larios:2002gq}, which shows the validity and reliability of our simple estimation method provided in this appendix.

Then we will follow almost the same procedure to approximately calculate the branching ratio of $Z\to \nu\bar{\nu}\gamma\gamma$ induced by our dimension-6 operator in Eq.~\eqref{Lagrangian} with the electron-neutrino current $J^{\nu_e}_\mu$. The corresponding Feynman diagrams are still represented by Fig.~\ref{FigZD} and the result is given by
\begin{eqnarray}\label{BrDim6}
	{\cal B} \left(Z\to \nu\bar{\nu}\gamma\gamma\right) \approx {\cal B} \left(Z\to \nu\bar{\nu}\right)\times 4 \times\frac{1}{\left(4\pi\right)^4}\times\beta^2 \left(\frac{E_{\nu,\gamma}}{M}\right)^4\,,
\end{eqnarray}   
which leads to the following upper bound on the effective coupling
\begin{eqnarray}
	\frac{\beta}{M^2} \lesssim  \frac{1.1 \times 10^3}{ \left({\rm TeV}\right)^{2}}\, \quad \mbox{at 95\% CL.}
\end{eqnarray}
In Sec.~\ref{neutrino}, we shall use this constraint in our numerical studies on the neutrino asymmetry explanation of the cosmic birefringence. }

\acknowledgments
This work is supported in part by  the National Natural Science Foundation of China (NSFC) (Grant No.~12005254 and No.~12147103) and the National Key Research and Development Program of China (Grant No.~2021YFC2203003 and No.~2020YFC2201501 ).


\appendix

\bibliography{CosBi}

\end{document}